\documentclass[12pt]{iopart}
\usepackage{epsfig,multirow}

\def\dddot#1{\mathinner{\buildrel\vbox{\kern5pt\hbox{...}}\over{#1}}}

\def\bs{\begin{subequations}}
\def\es{\end{subequations}}
\def\be{\begin{equation}}
\def\ee{\end{equation}}
\def\bq{\begin{eqnarray}}
\def\eq{\end{eqnarray}}
\def\beq{\begin{eqnarray*}}
\def\eeq{\end{eqnarray*}}
\def\ba{\begin{eqnarray}}
\def\ea{\end{eqnarray}}

\begin{document}

\title{New charged shear-free relativistic models with heat flux}

\author{ Y Nyonyi\dag ,
  S D Maharaj\dag  and K S Govinder\dag}
\address{\dag\ Astrophysics and Cosmology Research Unit,
School of Mathematics, Statistics and Computer Science,
 University of KwaZulu-Natal, Private Bag X54001,
 Durban 4000, South Africa \\ Email: maharaj@ukzn.ac.za
 }

\begin{abstract}
We study shear-free spherically symmetric relativistic gravitating fluids with heat flow and electric charge. The solution to the Einstein-Maxwell system is governed by the generalised pressure isotropy condition which contains a contribution from the electric field. This condition is a highly nonlinear partial differential equation. We analyse this master equation using Lie's group theoretic approach. The Lie symmetry generators that leave the equation invariant are found. The first generator is independent of the electromagnetic field. The second generator depends critically on the form of the charge, which is determined explicitly in general. We provide exact solutions to the gravitational potentials using the symmetries admitted by the equation. Our new exact solutions contain earlier results without charge. We show that other charged solutions, related to the Lie symmetries, may be generated using the algorithm of Deng. This leads to new classes of charged Deng models which are generalisations  of conformally flat metrics.
\end{abstract}

Keywords: Charged gravitating fluids; generalised Deng models; Einstein-Maxwell equations

PACS numbers: 02.30.Jr, 04.20.Jb, 04.40.Nr

\maketitle

\section{Introduction}
In this paper, we consider charged spherically symmetric gravitating fluids, in the presence of heat  flow, with vanishing shear which are important in the study of various cosmological and relativistic astrophysical bodies. It is necessary to solve the Einstein-Maxwell system of field equations to obtain exact solutions. Krasinski $\cite{k8}$ points out the importance of these solutions for modelling in structure formation, evolution of voids, gravitational collapse, inhomogeneous cosmologies and relativistic stellar processes. In these applications, heat flow and charge become important ingredients in building radiating and gravitating models. By studying shear-free models, we avail ourselves with a rather simpler avenue where we only need to provide solutions the generalised condition of pressure isotropy containing two metric functions. The resulting nonlinear equations with shear are much more difficult to analyse.

Heat flux is of great importance in relativistic astrophysical problems involving singularities in manifolds, gravitational collapse and black hole physics, among other applications as emphasised by Krasinski $\cite{k8}$. Such fluids have also been used in the study of relativistic stars that emit null radiation in the form of radial heat flow; a study made possible by Santos $\cite{sok}$ who showed that the interior spacetime must contain a nonzero heat flux to match with the pressure at the boundary with the exterior Vaidya spacetime. The notion of heat flow is manifested in many shear-free stellar models including the treatment of Wagh $\textit{et al}$ $\cite{wggmmm}$ who chose a barotropic equation of state and gave solutions to the Einstein field equations for a spherically symmetric spacetime. Maharaj and Govender $\cite{mg}$ and Misthry $\textit{et al}$ $\cite{mml}$, when studying radiating collapse with vanishing Weyl stresses, provided exact solutions to both the Einstein field equations and the junction conditions. Herrera $\textit{et al}$ $\cite{hdo}$ showed that analytic solutions can be obtained from the study of the field equations arising from radiating and collapsing spheres in the diffusion approximation. They showed that heat flow is a requirement in thermal evolution of the collapsing sphere modelled in causal thermodynamics. We note the recent general treatment of Thirukkanesh $\textit{et al}$ $\cite{trm1}$ for radiating spheres in the presence of shear in spherical symmetry.

In the cosmological setting, some of the earlier studies in which heat flow is an important component were carried out by Bergmann $\cite{b1}$, Maiti $\cite{m1}$, Modak $\cite{m2}$ and Sanyal and Ray $\cite{sr}$ in their quest to provide exact solutions. Deng $\cite{d1}$, using his general algorithm, regained earlier results and provided new classes of solutions. Msomi $\textit{et al}$ $\cite{mgm2}$ studied the same model and used Lie's group theoretic approach to provide a five-parameter family of transformations that mapped known solutions into new ones. They also obtained new classes of solutions using Lie infinitesimal generators. Later Msomi $\textit{et al}$ $\cite{mgm3}$ considered the problem in higher dimensions obtaining implicit solutions or reducing the fundamental equation to a Riccati equation. Also, Ivanov $\cite{ibv}$, using a compact formalism, simplified the condition of pressure isotropy and the condition for conformal flatness, and gave easily tractable versions of the junction condition for conformally flat and geodesic models. This approach has the advantage of yielding well known differential equations, amalgamates the results for static models, and places the time-dependent results of Msomi $\textit{et al}$ $\cite{mgm2, mgm3}$ in context. 

Stellar models in which charge is incorporated, so that the Einstein-Maxwell system is valid, have also been extensively studied. Komathiraj and Maharaj $\cite{km2}$ showed that by considering a linear equation of state, exact analytical solutions to the Einstein-Maxwell equations can be obtained that contains the Mak and Harko $\cite{mh}$ model. They obtained solutions that describe quark matter in the presence of an electromagnetic field. Other recent charged stellar models include the results of Komathiraj and Maharaj $\cite{km1}$, Lobo $\cite{l1}$, Maharaj and Thirukkanesh $\cite{mt}$, Sharma and Maharaj $\cite{sm1}$ and Thirukkanesh and Maharaj $\cite{tm2}$. Radiating stellar models where charge is incorporated have also been extensively studied by Chan $\cite{rc1, rc2}$ using numerical techniques. Recently, Pinheiro and Chan $\cite{pc1}$ performed a numerical analysis of a charged body undergoing gravitational collapse and showed that charge delays black hole formation and can even prevent collapse depending on the total mass-to-charge ratio. For varying spherically symmetric gravitational fields in cosmology, Kweyama $\textit{et al}$ $\cite{kgm1}$ found new parametric solutions to the Einstein-Maxwell system of field equations. Their approach was $\textit{ad hoc}$; a systematic approach using group theoretical techniques such as the Lie analysis may lead to new results. Govinder $\textit{et al}$ $\cite{glm}$, Kweyama $\textit{et al}$ $\cite{kgm2, kgm3}$, Leach and Maharaj $\cite{lm}$ and Msomi $\textit{et al}$ $\cite{mgm1}$ used Lie point symmetries to study the underlying nonlinear partial differential equations that arise in the study of gravitating fluids. They provided several families of solutions while generalising already known solutions. It is evident that there exists numerous physical applications to models in which heat flow and charge are incorporated.

Several techniques of obtaining solutions for gravitating fluids have been adopted over the years which yielded a variety of models. We intend to show that applying a group theoretic approach with Lie symmetries provides new insights for charged heat conducting models in the absence of shear. We present the field equations and obtain the defining master equation in $\S \ref{model}$.  We then obtain the underlying symmetries of the governing equation in $\S \ref{lie-analy-prob}$. This is a complex calculation and we provide all the relevant details. We use the first symmetry obtained to provide new solutions for an arbitrary form of charge in $\S \ref{usingx1}$. The gravitational potentials can be found explicitly. In the subsequent sections $\S \ref{usingx2} - \S \ref{f=ku}$, we show how the respective symmetries are used to reduce the order of the governing equation, while providing exact solutions to the gravitational potentials in some cases. The cases where reduction to quadrature is difficult to perform arise from the nonlinearity of the resultant equations. New Charged Deng solutions are obtained in $\S \ref{dengapp}$. A few concluding remarks follow in $\S \ref{concn}$.

\section{The model} \label{model}
We assume a spherically symmetric spacetime which satisfies the shear-free condition. Then the line element in Schwarzschild coordinates $(t,r,\theta ,\varphi)$ becomes
\be \label{line-element} \mathrm{d}s^{2} = -D^{2}\mathrm{d}t^{2} + \frac{1}{V^{2}} \left[ \mathrm{d}r^{2}+ r^{2} \left(\mathrm{d}\theta^{2} +  \sin^{2}\theta \mathrm{d}\varphi^{2} \right) \right] \ee
where $D=D(t,r)$ and $V=V(t,r)$ represent the gravitational potentials. We also define the energy momentum tensor for a charged matter distribution in a shear-free model to be of the form
\be \label{tab} T_{ab} = (\rho + p) U_{a} U_{b} + pg_{ab} + q_{a} U_{b} + q_{b} U_{a}  + E_{ab} \ee
where $\rho$ is the energy density, $p$ is the isotropic pressure, and $q_{a} = \left(0,q,0,0 \right)$ is the heat flux vector. These quantities are measured relative to a comoving four-velocity vector $U_{a} = \left(\frac{1}{D},0,0,0 \right)$ that is taken to be unit and timelike. The electromagnetic contribution $E_{ab}$ to the matter distribution is obtained from
\be E_{ab} = F_{ac} F^{c}{}_{b} - \frac{1}{4} g_{ab}F_{cd}F^{cd} \ee
where the Faraday tensor $$F_{ab} = A_{b;a} - A_{a;b}$$ is defined in terms of a four-potential $A_{a}= \left( \phi(t,r), 0,0,0 \right)$ with $\phi$ being the only nonzero component.

With the help of (\ref{line-element}) and (\ref{tab}), the Einstein-Maxwell field equations are given by
  \numparts
\begin{eqnarray} \label{EINMAXcomps}
\fl \rho=3\frac{V_{t}}{D^{2}V^{2}} + V^{2}\left[ 2\frac{V_{rr}}{V} - 3\frac{V_{r}^{2}}{V^{2}} + 4\frac{V_{r}}{rV} \right] -\frac{V^{2}}{2D^{2}} \phi^{2}_{r} \label{EINMAXcomp1} \\
\fl p = \frac{1}{D^{2}} \left[ 2\frac{V_{tt}}{V} -2 \frac{D_{t} V_{t}}{DV} -5\frac{V_{t}^{2}}{V^{2}} \right] + V_{r}^{2} - 2\frac{VV_{r}}{r} - 2\frac{VD_{r}V_{r}}{D} +2\frac{D_{r}V^{2}}{rD} +\frac{1}{2}\frac{V^{2}}{D^{2}} \phi^{2}_{r} \label{EINMAXcomp2}\\
\fl p = \frac{1}{D^{2}} \left[ 2\frac{V_{tt}}{V} -2 \frac{D_{t} V_{t}}{DV} -5\frac{V_{t}^{2}}{V^{2}} \right] -VV_{rr} + V_{r}^{2} -\frac{VV_{r}}{r} + \frac{V^{2}D_{r}}{rD} + \frac{V^{2}D_{rr}}{D} - \frac{1}{2}\frac{V^{2}}{D^{2}} \phi^{2}_{r} \label{EINMAXcomp3} \\
\fl q = -2\frac{V^{2}}{D}\left[ \frac{V_{tr}}{V} - \frac{V_{t}V_{r}}{V^{2}} -\frac{D_{r}V_{t}}{DV} \right] \label{EINMAXcomp4}  \\
\fl \sigma = \frac{V^{2}}{D} \left[ \phi_{rr} + \left( \frac{2}{r} - \frac{V_{r}}{V} - \frac{D_{r}}{D} \right) \phi_{r} \right] \label{EINMAXcomp5} \\
\fl 0 = -\frac{V^{2}}{D^{2}} \left[ \phi_{rt} - \left( \frac{V_{t}}{V} + \frac{D_{t}}{D} \right) \phi_{r} \right] \label{EINMAXcomp6}
\end{eqnarray}
\endnumparts
where $\sigma$ is the proper charge density. On integrating (\ref{EINMAXcomp6}), we obtain
\be \label{phi-r} \phi_{r} = VDF(r) \ee
where $F(r)$ is an arbitrary function. Equating (\ref{EINMAXcomp2}) and (\ref{EINMAXcomp3}) gives the generalised pressure isotropy condition
\be \label{comp2ncomp3}- \frac{VV_{r}}{r} - 2\frac{VD_{r}V_{r}}{D} + \frac{D_{r}V^{2}}{rD} + VV_{rr} - \frac{V^{2}D_{rr}}{D} + \frac{V^{2}}{D^{2}} \phi^{2}_{r}=0  \ee

The system (\ref{EINMAXcomps})  is completely solved if we can find functions $\phi$, $V$ and $D$ that satisfy (\ref{comp2ncomp3}). Therefore (\ref{comp2ncomp3}) is the fundamental equation governing the evolution of a shear-free, heat conducting gravitating fluid. The generalised pressure isotropy condition (\ref{comp2ncomp3}) simplifies to
\be \label{master-charged} 4uVD_{uu} +8uV_{u}D_{u} - 4uDV_{uu} - V^{2} F(u) = 0 \ee
with $u=r^{2}$ and $F(u)$ is arbitrary. In this paper, we seek to provide solutions to this master equation (\ref{master-charged}) using Lie's group theoretic approach. Note that in the absence of charge (\ref{master-charged}) becomes
\be \label{master-uncharged} VD_{uu} +2V_{u}D_{u} - DV_{uu} = 0 \ee
which was studied by Msomi $\textit{et al}$ $\cite{mgm1}$.

\section{Lie analysis of the problem} \label{lie-analy-prob}
For the problem at hand, we seek to determine a one-parameter ($\varepsilon$) Lie group of transformations
\numparts
\begin{eqnarray}
\tilde{u} = \mathrm{f}(u,V,D;\varepsilon )\label{global-infs-udv1}\\
\tilde{D} = \mathrm{g}(u,V,D;\varepsilon ) \label{global-infs-udv2}\\
\tilde{V} = \mathrm{h}(u,V,D;\varepsilon ) \label{global-infs-udv3}
\end{eqnarray}
\endnumparts
that leave the solutions of (\ref{master-charged}) invariant. Due to the complexity involved in obtaining the transformations directly, we consider the infinitesimal forms
\numparts
\begin{eqnarray}
\tilde{u} = u + \varepsilon \xi(u,V,D) + O(\varepsilon^2) \label{infs-udv1}\\
\tilde{D} = D + \varepsilon \eta^{1}(u,V,D) + O(\varepsilon^2) \label{infs-udv2}\\ 
\tilde{V} = V + \varepsilon \eta^{2}(u,V,D) + O(\varepsilon^2)\label{infs-udv3}
\end{eqnarray}
\endnumparts
with symmetry generator given by 
\be \label{Inf} X = \xi \frac{\partial}{\partial u} + \eta^1  \frac{\partial}{\partial D} + \eta^2  \frac{\partial}{\partial V} \ee
We can then infer from (\ref{infs-udv1})-(\ref{infs-udv3}), to regain the global form of the transformations (\ref{global-infs-udv1})-(\ref{global-infs-udv3}), that we need to solve 
\numparts
\begin{eqnarray}
\frac{\mathrm{d}\tilde{u}}{\mathrm{d} \varepsilon} =\xi(u,V,D) \\
\frac{\mathrm{d}\tilde{D}}{\mathrm{d} \varepsilon} = \eta^{1}(u,V,D) \\
\frac{\mathrm{d}\tilde{V}}{\mathrm{d} \varepsilon} = \eta^{2}(u,V,D)
\end{eqnarray}
\endnumparts
subject to 
\be  \tilde{u} \vert_{\varepsilon = 0} = u, \qquad \tilde{D} \vert_{\varepsilon = 0} = D, \qquad \tilde{V} \vert_{\varepsilon = 0} = V \ee 
For a detailed review of these ideas, the reader is referred to Bluman and Anco $\cite{ba1}$, Bluman and Kumei $\cite{bk1}$ and Olver $\cite{o1,o2}$.

Due to the complexity of the calculations, we provide as much relevant detail as possible. It is important to note that both $D$ and $V$ are functions of $u$ and $t$, but $t$ does not appear explicitly in equation (\ref{master-charged}). As a result we can treat (\ref{master-charged}) as a second order nonlinear ordinary differential equation only in $u$. However, we will ultimately let the constants of integration become functions of $t$. For simplicity, we label the left hand part of our master equation (\ref{master-charged}) as $K$. We require
\be X^{[2]}K|_{K=0} = 0 \ee
which yields
\be \xi = C^{0}(u) \label{xi222}\ee
\be \eta^{1} = c_{1}D \label{eta111} \ee
\be \eta^{2} = \left( c_{1} + c_{2}+ \frac{1}{2}C^{0}_{u} \right) V \label{eta222} \ee
with $F(u)$ satisfying 
\be \label{sys2aa} 2VDC^{0}_{uuu} + V^{2} \left[ F \left( -\frac{C^{0}}{u} + \frac{5}{2}C^{0}_{u} +c_{2}\right)+C^{0}F^{\prime} \right] = 0 \ee
For arbitrary $F$,  (\ref{sys2aa}) can only be satisfied if 
\numparts
\begin{eqnarray} 
 C^{0}_{uuu} = 0 \label{C01} \\ 
 -\frac{C^{0}}{u} + \frac{5}{2}C^{0}_{u} +c_{2} = 0 \label{fffffff} \\
 C^{0} = 0 \label{fpfpfp} 
\end{eqnarray}
\endnumparts
By inspection, we can deduce from equations (\ref{C01})-(\ref{fpfpfp}) that
\be \label{co-arbit-f} C^{0}(u) = 0 \ee
and 
\be \label{c2-arbit-f} c_{2} = 0 \ee
Using (\ref{xi222})-(\ref{eta222}) and (\ref{co-arbit-f})-(\ref{c2-arbit-f}), and making the necessary substitutions, we obtain the coefficient functions for the symmetry generator, when $F(u)$ is arbitrary, as
\be \xi(u) = 0 \ee 
\be \eta^{1}(D) = c_{1} D \ee
\be \eta^{2}(V) = c_{1} V \ee
From the above coefficient functions, we obtain
\be X_{1} = D \partial_{D} + V \partial_{V} \label{x1} \ee
as the sole symmetry in the case of arbitrary $F(u)$. It is indeed remarkable that this symmetry exists without placing any restriction on $F(u)$.

We now take (\ref{sys2aa}) to be a restriction on $F$. As $C^{0}$ and $F$ are functions of $u$, this implies that both 
\be \label{C00} C^{0}_{uuu} = 0 \ee 
and 
\be \label{fu} F \left( -\frac{C^{0}}{u} + \frac{5}{2}C^{0}_{u} +c_{2}\right) + C^{0}F^{\prime} = 0 \ee
must hold. Solving equation (\ref{C00}) gives 
\be \label{C001} C^{0} = c_{3}u^{2} + c_{4}u + c_{5} \ee
Using equation (\ref{C001}) we solve (\ref{fu}) to obtain
\be  \fl \label{fgeneral} F(u) = \frac{uc_{6}}{(c_{3}u^{2} + c_{4}u + c_{5})^{5/2}}  \exp \left[ \frac{2 c_{2}}{\sqrt{-c_{4} + 4c_{3} c_{5}}} \arctan \left(\frac{c_{4}+2c_{3}u}{\sqrt{-c_{4} + 4c_{3}c_{5}}} \right) \right]  \ee
where $c_{6}$ is a constant of integration. From equations (\ref{C001})-(\ref{fgeneral}), we see that in addition to $X_{1}$, (\ref{master-charged}) admits another symmetry 
\be X_{2} = c_{1}X_{1} + \left(c_{3}u^{2} + c_{4}u + c_{5} \right) \partial_{u} + \left(c_{2} + c_{3}u + \frac{c_{4}}{2}\right)V \partial_{V} \label{x2} \ee
dictated by the form of $F(u)$. It is important to observe that the quantity $F(u)$ arises because  of the presence of charge. The symmetry $X_2$ is intimately related to the form of the electromagnetic field. It is remarkable that the electric field, through (\ref{fgeneral}), can be explicitly found in general when the symmetry generator $X_2$ exists.

Equation (\ref{fgeneral}) gives the most general form of $F(u)$ for which $X_{2}$ is the associated symmetry. We note that $F(u)$ depends on the arbitrary constants $c_{2} - c_{6}$. Of these, four, $c_{2} - c_{5}$, appear in the symmetry itself while $c_{6}$ is just a scaling constant. Clearly, choices for $c_{2} - c_{5}$ will produce simpler forms of $F(u)$ which will admit reduced forms of $X_{2}$ as a symmetry (in addition to $X_{1}$). In Table $1$, we list  the relevant simpler forms for $F(u)$ and the corresponding form of $X_{2}$ for all the relevant choices of $c_{2} - c_{5}$. (Note that, in all cases, we relabel our scaling constant for $F(u)$ to be $k$.) In some cases, the simpler form of $F(u)$ causes the original equation (\ref{master-charged}), to admit additional symmetries. These are listed in the third column of Table $1$. For the case $F=0$, a full analysis was performed by Msomi $\textit{et al}$ $\cite{mgm2}$, and we do not repeat their results here. We obtained extra symmetries for $c_{2}$, $c_{3}$ and $c_{5}$ only.  This is summarised in Table $1$. The only other case of interest is when $c_{2} = 0$ (with all other constants being nonzero) as reported in Table $1$.

\begin{table}
\caption{\label{syms-fu}Symmetries associated with different forms of $F(u)$.} 
\begin{indented}
\lineup
\item[]\begin{tabular}{@{}*{3}{l}}
\br                              
Symmetry generator &Form of $F(u)$ & Extra symmetries\cr 
\mr
\multirow{4}*{$X_{2,1} = V \partial_{V}$}  & \multirow{4}*{$0$} & $X_{3} =\partial_{u}$,  \cr
   &      &$X_{4} = u\partial_{u} $, \cr
   &      &$X_{5}=D\partial_{D} $, \cr
   &      &$X_{6} = u^{2} \partial_{u} + uV \partial_{V}$  \cr
$X_{2,2} = u^2\partial_{u} +uV\partial_{V}$  & $k u^{-4}$ & $X_{3} = \partial_{u} +3V\partial_{V}$  \cr
$X_{2,3} = u\partial_{u} +\frac{1}{2} V\partial_{V}$  & $ ku^{-\frac{3}{2}}$ & None  \cr
$X_{2,4} = \partial_{u}$  & $ k u$ & $X_{3} = -\partial_{u} +2V\partial_{V}$ \cr 
 $X_{2,5} = \left(c_{3}u^{2} + c_{4}u + c_{5} \right)\partial_{u}$ &  \multirow{2}*{$ k(c_{3}u^{2} + c_{4}u + c_{5})^{-5/2}$} & \multirow{2}*{None } \cr 
 $\qquad+ \left(c_{3}u + \frac{c_{4}}{2} \right)\partial_{V} $ &  &  \cr
\br
\end{tabular}
\end{indented}
\end{table}

\section{New solutions using symmetries}
Usually, after obtaining the symmetries of a differential equation, we use the associated differential invariants to determine the solution(s) of the equation. For all cases, we were able to reduce the order of our master equation. However we were not always able to solve the reduced equation. In particular no solutions were possible for the symmetries $X_{2,5}$ and $X_{2}$. We discuss these below.
\subsection{Arbitrary $F$} \label{usingx1}
Due to the arbitrary nature of $F$, our master equation can be modified to be
\be \label{arbtf} VD_{uu} +2V_{u}D_{u} - DV_{uu} - V^{2} \frac{F(u)}{4u}  = 0 \ee
We obtain the invariants of the generator
\be X_{1} = D \partial_{D} + V \partial_{V} \ee
by taking its first extension. The associated Lagrange's system becomes
\be \frac{\mathrm{d}u}{0} = \frac{\mathrm{d}D}{D} = \frac{\mathrm{d}V}{V} = \frac{\mathrm{d}D^{\prime}}{D^{\prime}} = \frac{\mathrm{d}V^{\prime}}{V^{\prime}} \ee
We obtain the invariants of the system as
\numparts
\begin{eqnarray*}
 p = u \\
q(p) = \frac{V}{D} \\ 
r(p) = \frac{D^{\prime}}{D} \\
s(p) = \frac{V^{\prime}}{D}
\end{eqnarray*}
\endnumparts
However, for our purposes, we only use $p$, $q$ and $r$. Invoking these differential invariants, (\ref{arbtf}) reduces to
\be q^{\prime \prime} = -  q^{2} \frac{F(p)}{4p} + 2qr^{2} \ee
which can be written as
\be \label{rpqu} r = \pm \sqrt{\frac{q^{\prime \prime}}{2q} + q\frac{F(p)}{8p}} \ee
or
\be \frac{D^{\prime}}{D} = \pm \sqrt{\frac{q^{\prime \prime}}{2q} + q \frac{F(p)}{8p}}\ee
On integrating both sides we have
\be \label{soln1} D = \exp \left[\pm C \int  \sqrt{\frac{W^{\prime \prime}}{2W} +  W \frac{F(u)}{8u}} \mathrm{d}u \right] \ee
where $C$ is a constant of integration.

From solution (\ref{soln1}), we can see that whenever we are given any ratio of the gravitational potentials $W=\frac{V}{D}$, and an arbitrary function $F(u)$ representing charge, we can explicitly  obtain the exact expression of the potentials. This is a new result that to the best of our knowledge has not been obtained before. We observe that when we set $F(u)=0$ in (\ref{soln1}), we obtain 
\be \label{solnmmg} D = \exp \left[\pm C \int  \sqrt{\frac{W^{\prime \prime}}{2W}} \mathrm{d}u \right] \ee
This is the uncharged solution of Msomi $\textit{et al}$ $\cite{mgm2}$. Thus (\ref{soln1}) is a charged generalisation of their solution.

\subsection{$F = 0 $} \label{usingx2}
For this particular form of $F$, our master equation takes the form
\be \label{f=0} VD_{uu} +2V_{u}D_{u} - DV_{uu} = 0 \ee
By taking the first extension of 
\be X_{2,1} = V\partial_{V } \ee 
the associated Lagrange's system becomes
 \be \frac{\mathrm{d}u}{0} = \frac{\mathrm{d}D}{0} = \frac{\mathrm{d}V}{V} = \frac{\mathrm{d}D^{\prime}}{0} = \frac{\mathrm{d}V^{\prime}}{V^{\prime}} \ee
The corresponding invariants become
\numparts
\begin{eqnarray*}
p = u \\
q(p) = D \\
r(p) = D^{\prime}  \\ 
s(p) = \frac{V^{\prime}}{V}
\end{eqnarray*}
\endnumparts
When we use a partial set of invariants $p$, $q(p)$ and $s(p)$, (\ref{f=0}) reduces to
\be \label{f01} s_{p} = \frac{q_{pp}}{q} + 2\frac{q_{p}}{q} s - s^{2} \ee
which is a Riccati equation in $s$. It is not possible to make further progress with (\ref{f01}).

However, if we include $r(p)$ and consider the full set of invariants, (\ref{f=0}) reduces to
\be \label{x2reduced2} r_{p} + 2sr = \left( s_{p} + s^2 \right)q \ee
Equation (\ref{x2reduced2}), being a first order differential equation in $r$, can easily be reduced to quadrature to give
\be \label{x2reduced3} r(p) = e^{-2s} \int \left(s_{p} +s^{2} \right)q e^{2s}  \mathrm{d}p \ee
From our invariants it easily follows that 
\be D = \int \left(e^{-2 ( V^{\prime}/V)} \int D \left[ \frac{\mathrm{d}(V^{\prime}/V)}{\mathrm{d}u} + \left( \frac{ V^{\prime} }{V} \right)^{2} \right]  e^{2(V^{\prime}/V) } \mathrm{d}u \right) \mathrm{d}u \ee
This result was first established by Msomi $\textit{et al}$ $\cite{mgm2}$ and a comprehensive study of the uncharged case produced five symmetries as already indicated above. They provided the complete analysis of the uncharged model and we do not intend to reproduce their results herein.

\subsection{$F = k u^{-4}$} \label{f=ku-4}
Our master equation becomes
\be \label{fu=ku4} 4uVD_{uu} +8uV_{u}D_{u} - 4uDV_{uu} - V^{2}k u^{-4} = 0 \ee
For this case, we obtain two extra symmetries associated with the form of $F$ highlighted above. We carry out reductions using these symmetries separately with the hope of obtaining new solutions.
\subsubsection{Generator $X_{2,2}$}\label{usingx2,2}
By taking the first extension of
\be  X_{2,2} = u^2 \partial_{u} +uV\partial_{V} \ee
we obtain the Lagrange's system from which we deduce the invariants to be
\numparts
\begin{eqnarray*}
p    = D  \\
q(p) = \frac{V}{u}   \\
r(p) = u^{2} D^{\prime}  \\
s(p) = u\left(q(p)- V^{\prime} \right)
\end{eqnarray*}
\endnumparts
Using the invariants $p$, $q$ and $r$, equation (\ref{fu=ku4}) reduces to
\be 4rqr_{p} + 8r^{2}q_{p} - 4(r r_{p} q_{p} + r^{2} q_{pp})p - kq^{2} =0 \ee
or
\be \label{x3reduced2} \frac{1}{2} (r^{2})_{p} + \left(\frac{ 2q_{p} - pq_{pp}}{q- pq_{p}} \right)r^{2} -\left( \frac{(k/4) q^{2}}{q- pq_{p}} \right) = 0 \ee
Equation (\ref{x3reduced2}) is a first order differential equation in $r^2$ which when solved gives
\be \label{x3reduced3} r(p) = \sqrt{\frac{Y + e^{X(p+ A)} }{X}} \ee
where
\be Y = \frac{(k/2) q^{2}}{q- pq_{p}}, \qquad  X = 2 \left(\frac{ 2q_{p} - pq_{pp}}{q- pq_{p}} \right)  \ee
and $A$ is a constant of integration.

By taking the invariants into consideration, (\ref{x3reduced3}) is reduced to quadrature to give
\be \int \left(\frac{X}{e^{X(D + A)} +Y}\right)^{1/2} \mathrm{d} D = -\frac{1}{u} + B \ee
where 
\be X = 2\frac{2 \frac{\mathrm{d}(V/u)}{\mathrm{d}D} - D \frac{{\mathrm{d}}^2(V/u)}{{\mathrm{d}D}^2}     }{(V/u)-  D \frac{\mathrm{d}(V/u)}{\mathrm{d} D}}, \qquad
Y =\frac{(k/4) (V/u)^{2}}{(V/u)-  D \frac{\mathrm{d}(V/u)}{\mathrm{d} D}} \ee
and $B$ is a constant of integration.
\subsubsection{Generator $X_{3}$} \label{usingx3-1}
This is the second extra symmetry associated with (\ref{fu=ku4}). By taking the first extension of 
\be X_{3}= u\partial_{u} +3V\partial_{V} \ee
we obtain the corresponding Lagrange's system from which the invariants become
\numparts
\begin{eqnarray*}
p    = D  \\
q(p) = \frac{\sqrt[3]{ V}}{u} \\
r(p) = u D^{\prime} \\
s(p) = \frac{\sqrt{V^{\prime}}}{u}
\end{eqnarray*}
\endnumparts
The first derivatives of $r(p)$ and $s(p)$ with respect to $p$ enable us to obtain expressions for $D^{\prime \prime} $ and $V^{\prime \prime}$ respectively. Thus we are able to transform equation (\ref{fu=ku4}) to
\be \label{x31reduced} ss_{p} + s^2 \left( \frac{2}{r} - \frac{1}{p} \right) + \frac{q^3}{p} \left(1 - r_{p} + \frac{kq^3}{4r} \right) = 0\ee
Equation (\ref{x31reduced}) is a first order differential equation in $s^2$, which when solved gives
\be s(p) =  \pm \sqrt{ \frac{ \frac{q^3}{p}\left( 1-r_{p} + k\frac{q^3}{4r} \right)+ e^{-4(1/r - 1/p)(p-C)}}{\frac{2}{p} \left( \frac{p}{r} - 1 \right)} }\ee
where $C$ is a constant of integration.

By taking the invariants into consideration, we obtain explicitly the exact solution of one of the potentials as
\be \int \frac{\mathrm{d} V}{\frac{u^{3} D D^{\prime}}{2(D- uD^{\prime}} \left[ \frac{V}{Du^{3}} \left( \frac{uD^{\prime \prime }}{D} - \frac{kV}{4u^{4}D^{\prime}} \right) + e^{\frac{4(uD^{\prime} - D)(D-C)}{uD^{\prime}D}} \right]} = \frac{u^3}{3} + E \ee
where $E$ is a constant of integration. 
 
\subsection{$F(u) = ku^{-\frac{3}{2}}$} \label{fu=ku-3/2}

For this particular form of $F$, the master equation becomes
\be \label{fu=ku-3/2} 4uVD_{uu} +8uV_{u}D_{u} - 4uDV_{uu} - V^{2}ku^{-\frac{3}{2}} = 0 \ee
By taking the first prolongation of the associated generator 
\be X_{2,3} = u\partial_{u} +\frac{1}{2} V\partial_{V} \ee
we obtain the corresponding Lagrange's system from which the invariants become
\numparts
\begin{eqnarray*}
p    = D  \\
q(p) = \frac{V^{2}}{u}   \\
r(p) = u D^{\prime}  \\
s(p) = u {V^{\prime}}^{2}
\end{eqnarray*}
\endnumparts
Making use of the first three invariants $p$, $q$ and $r$, equation (\ref{fu=ku-3/2}) reduces to 
\be 4rqr_{p} + 4r^{2}q_{p} - p \left( 2r^{2}q_{pp} - r^{2}q_{p} + 2rr_{p}q_{p}- q \right) -  kq^{\frac{3}{2}} = 0 \ee
or
\be \label{reduced42} (r^2)_{p} + \frac{4q_{p} - p(2q_{pp} - q_{p}^{2})}{2q-pq_{q}}r^{2} + \frac{pq-kq^{3/2}}{2q-pq_{q}} = 0 \ee
Equation (\ref{reduced42}) is a first order differential equation in $r^2$ which can be solved to obtain
\be \label{reduced43} r(p) = \sqrt{\frac{e^{Y(p+A)}-Z}{Y}} \ee
where
\begin{equation*}  Y = \frac{4q_p - p(2q_{pp}-q^{2}_{p})}{2q - pq_{p}}, \quad Z = \frac{pq-kq^{(3/2)}}{2q - pq_{p}}  \end{equation*}
and $A$ is a constant of integration.

Using the invariants, we can provide the explicit solution to (\ref{reduced43}) as
\be  \label{reduced44}\int \left( \frac{Y}{e^{Y(p+A)}-Z} \right)^{1/2} \quad \mathrm{d} D = \ln u + B \ee
where
\begin{equation*}  Y = \frac{4\frac{\mathrm{d}(V^2/u)}{\mathrm{d} D} - D \left[ \frac{{\mathrm{d}}^{2}(V^2/u)}{{\mathrm{d} D}^2} - \left(\frac{\mathrm{d}(V^2/u)}{\mathrm{d} D}\right)^{2} \right]}{2(V^2/u) - D \frac{\mathrm{d}(V^2/u)}{\mathrm{d} D} }, \quad Z = \frac{D(V^2/u)-k(V^2/u)^{(3/2)}}{2(V^2/u) - D \frac{\mathrm{d}(V^2/u)}{\mathrm{d} D}} \nonumber \end{equation*}
and $B$ is a constant of integration. 

\subsection{$F= ku$} \label{f=ku}

The master equation to be reduced is of the form
\be \label{fu=ku} 4uVD_{uu} +8uV_{u}D_{u} - 4uDV_{uu} - V^{2}ku = 0 \ee
We use the associated generator
\be X_{2,4} = \partial_{u}, \ee
so that we can obtain other forms of the potentials without having to make any restrictions on how the potentials relate initially. We obtain the invariants of the generator above from its Lagrange's system after taking its first prolongation. The invariants become
\numparts
\begin{eqnarray*}
 p = D \\
q(p) = V \\ 
r(p) = D^{\prime} \\
s(p) = V^{\prime} 
\end{eqnarray*}
\endnumparts
We only use $p$, $q$ and $r$ for our purposes. Invoking these differential invariants, (\ref{fu=ku}) transforms to
\be qrr_{p} + 2r^{2}q_{p} - p \left( q_{pp}r^{2} + q_{p}r_{p}r \right) - Aq^{2} = 0  \ee
which can be written as
\be  \label{master2} r_{p} \left( q - pq_{p} \right) + r \left(2q_{p} - pq_{pp}  \right) -r^{-1} Aq^{2} = 0  \ee
A closer inspection of (\ref{master2}) reveals that it is indeed a Bernoulli equation of the form
\be  \label{master3} r_{p} + P(p,q)r - r^{-1}Q(p,q) = 0  \ee
with 
\begin{equation*}
P(p,q) = \frac{2q_{p} - pq_{pp}}{q - pq_{p}}, \quad Q(p,q) = \frac{Aq^{2}}{q - pq_{p}}
\end{equation*}
The solution to (\ref{master3}) becomes
\be r = \pm \sqrt{2 e^{-2 \int P(p,q) \mathrm{d}p} \int Q(p,q) e^{2 \int P(p,q) \mathrm{d}p}      \mathrm{d}p } \ee
or 
\be D^{\prime} = \pm \sqrt{2 e^{-2 \int P(p,q) \mathrm{d}p} \int Q(p,q) e^{2 \int P(p,q) \mathrm{d}p} \mathrm{d}p } \ee

On integrating both sides we have
 \be \int \frac{\mathrm{d} D}{\pm \sqrt{ 2 e^{-2 \int P(D,V) \mathrm{d}D} \int Q(D,V) e^{2 \int P(D,V) \mathrm{d}D} \mathrm{d}D}} =  u + C\ee
 where 
\begin{equation*} 
P(D,V) = \frac{2(\mathrm{d}V/\mathrm{d}D) - D (\mathrm{d}^{2} V/\mathrm{d}D^2)}{V - D(\mathrm{d}V/\mathrm{d}D)} \quad Q(D,V) = \frac{AV^{2}}{V -D(\mathrm{d}V/\mathrm{d}D)} \end{equation*}
and $C$ is a constant of integration.

We see that without prescribing any restriction on the relationship between the gravitational potentials, we can explicitly give the exact form of the potentials if the function $F(u)$ is linear in $u$.

\section{Charged Deng solutions} \label{dengapp}
Deng $\cite{d1}$ proposed a general algorithm, which can be applied indefinitely, by alternating between choices of $D$ and $V$ for uncharged matter ($F(u)=0$). This was possible as the equation could be treated as linear in $D$ or $V$. He reproduced several classes of solutions to the uncharged shear-free heat conducting fluids that were initially obtained by Bergmann $\cite{b1}$, Maiti $\cite{m1}$, Modak $\cite{m2}$ and Sanyal and Ray $\cite{sr}$ as well as generating new solutions. The Deng approach is powerful as all known uncharged models with heat flux can be regained from this general class of solutions. In the general case of $F(u) \neq 0$, if we choose $D$, (\ref{master-charged}) is a nonlinear equation in $V$ and is difficult to solve in general. However, if we choose forms for $V$, the resulting differential equation in $D$ is linear and, in principle, can be solved.

We illustrate this approach by taking some of Deng's $\cite{d1}$ seed solutions for $V$. The first solution we utilise is $V=1$. In this case, we can completely solve (\ref{master-charged}) for $D$ with $F(u)$ arbitrary. To link these results with those obtained from the symmetry analysis, we can also derive solutions for $D$ corresponding to the different group-invariant forms of $F(u)$. All these results are contained in Table $2$. Note that these results are charged generalisations of the Deng $\cite{d1}$ solutions with $V=1$. When the charge vanishes we regain $D= au+b$ and so
\be \label{lineelm1} \mathrm{d}s^{2} = - \left(au+b\right)^2 \mathrm{d}t^{2} + \mathrm{d}r^{2} \left(\mathrm{d}\theta^{2} + \sin^{2}\theta \mathrm{d}\varphi^{2} \right) \ee
(first obtained by Bergmann $\cite{b1}$).

We next take $V=au+b$ and again solve of $D$. The results for arbitrary $F(u)$ and group-invariant forms of $F(u)$ are given in Table $3$. This yields another new class of charged  solutions that generalise Deng's $\cite{d1}$ results. When the charge vanishes we regain $D= \frac{cu+d}{au+b}$ and so
\be \label{lineelm2} \mathrm{d}s^{2} = - \left(\frac{cu+d}{au+b}\right)^2 \mathrm{d}t^{2} + \left(au+b\right)^2 \left[\mathrm{d}r^{2} \left(\mathrm{d}\theta^{2} + \sin^{2}\theta \mathrm{d}\varphi^{2} \right)\right] \ee
The metric (\ref{lineelm2}) is the most general shear-free spherically symmetric form that is conformally flat, and was obtained by Modak $\cite{m2}$ and Sanyal and Ray $\cite{sr}$ independently. Thus we have obtained a new family of charged  models with heat flux that have vanishing Weyl tensor when the electric field vanishes.

This approach can be continued for different chosen forms of $V$. As we only need to solve a linear equation in $D$, the solution is usually obtained using standard techniques.

\begin{table}
\caption{\label{tab2}$V=1$} 

\begin{indented}
\lineup
\item[]\begin{tabular}{@{}*{3}{l}}
\br                              
Symmetry generator & $F(u)$ & $D(u)$ \cr 
\mr
$X_{1} $ &  Arbitrary    & $au+b +\int^{u} \int^{t} \frac{F(s)}{4s}$ $\mathrm{d}s \mathrm{d}t $ \cr
$X_{2,4}$ &$ku$  & $au+b + \frac{ku^2}{8}$  \cr 
$X_{2,3}$ &$ku^{-3/2}$ & $au+b + \frac{k}{3\sqrt{u}}$  \cr 
$X_{2,2}$ &$ku^{-4}$ &$au+b + \frac{k}{48u^{3}}$    \cr 
$X_{2,5}$  &$ku \left( c_{3}u^2 +c_{4}u + c_{5} \right)^{-5/2}$ & $au+b + \frac{(8uc_{3}(uc_{3}+c_{4})+ c_{4}^{2} +4c_{3}c_{5})k}{3(c_{4}^{2} - 4c_{3}c_{5})\sqrt{c_{3}u^2 +c_{4}u + c_{5}}}$    \cr
\br
\end{tabular}
\end{indented}
\end{table}

\begin{table}
\caption{\label{tab3}$V=au+b$} 

\begin{indented}
\lineup
\item[]\begin{tabular}{@{}*{3}{l}}
\br                              
Symmetry generator & $F(u)$ & $D(u)$ \cr 
\mr
\multirow{2}*{$X_{1} $} & \multirow{2}*{ Arbitrary}   & $\frac{cu+d}{au+b} $ \cr
  &     &   $ + \int^{u} \int^{t}\frac{as+b}{(at+b)^2}\frac{F(s)}{4s}\left( b^{2}+2abs+ a^{2}s^{2}\right) \mathrm{d}s\mathrm{d}t $ \cr
$X_{2,4}$ & $ku$  & $\frac{cu+d}{au+b} + \frac{ku^2}{8}\left(\frac{a^2u^2 + 4abu+ 6b^2}{au+b}\right)$ \cr
$X_{2,3}$ & $ku^{-3/2}$ & $\frac{cu+d}{au+b} + \frac{k}{3\sqrt{u}}\left(\frac{a^2u^2 - 6abu + b^2}{au+b}\right)$ \cr
$X_{2,2}$ & $ku^{-4}$ &$\frac{cu+d}{au+b} + \frac{k}{48u^{3}}\left( \frac{6a^2u^2 +4abu+b^2}{au+b} \right)$ \cr
\multirow{2}*{$X_{2,5}$} & \multirow{2}*{$ku \left( c_{3}u^2 +c_{4}u + c_{5} \right)^{-5/2}$} & $\frac{cu+d}{au+b} + \frac{k(au+b)}{3\sqrt{c_{3}u^2 +c_{4}u + c_{5}}} $ \cr
                 &      &  $  + \frac{ 8k\sqrt{c_{3}u^2 +c_{4}u + c_{5}}(b^2 c_{3} - abc_{4} + a^2 c_{5})}{3(au+b)(c_{4}^{2} - 4c_{3}c_{5})}$ \cr
\br
\end{tabular}
\end{indented}
\end{table}

\section{Conclusion} \label{concn}
We have obtained new exact solutions to the Einstein-Maxwell system of charged relativistic fluids in the presence of heat flux. Solutions to this highly nonlinear system were obtained by essentially solving the generalised pressure isotropy condition. A suitable transformation reduced the master equation to a second order nonlinear differential equation. The Lie symmetry generators for this master equation were found. Importantly, the first Lie generator does not depend on the electromagnetic field. The second Lie generator arose because of specific forms of the electric field; the electric charge for the Lie generator was found explicitly in general. In some cases additional symmetries are possible depending on the specific forms of the charge $F(u)$; these are identified in Table $1$. Solutions of the Einstein-Maxwell system were found corresponding to particular Lie symmetry generators.  In the case of arbitrary $F(u)$, we were able to give an explicit relationship between the metric functions $D$ and $V$ (via $W$). For any chosen form of $W$, one could find $D$ and $V$ explicitly. This approach is a generalisation of the Msomi $\textit{et al}$ $\cite{mgm1}$ method for the uncharged case. We believe that these results are new and have not been published before.

We also modified the method of Deng $\cite{d1}$ to obtain two new families of charged heat conducting relativistic fluids. Table $2$ provides charged generalisations of the Bergmann $\cite{b1}$ models. Table $3$ contains charged generalisations of shear-free conformally flat models which include the presence of the electric field. Both new families of solutions to the Einstein-Maxwell system are characterised geometrically by the infinitesimal Lie symmetry generator $X_{1}$. What is particularly remarkable about these new families of solutions is that they can be determined for arbitrary charge. Thus once a physically reasonable or observed charge is determined, the spacetime can be generated immediately.

~\\
{\bf Acknowledgements}\\
YN, KSG and SDM wish to thank the National Research Foundation and the University of KwaZulu-Natal for support. SDM acknowledges that this work is based on research supported by the South African Research Chair Initiative of the Department of Science and Technology and the National Research Foundation. 
~\\

\end{document}